\documentclass[]{mn2e}  
\usepackage{mnras_cite}  
\newcommand{\nc}{\newcommand}    
    
\nc{\be}[1]{\begin{equation}\mbox{$\label{#1}$}}    
\nc{\bea}[1]{\begin{eqnarray} \mbox{$\label{#1}$}}    
\nc{\Section}[2]{\section{#2}\label{#1}}    
\nc{\Bibitem}[1]{\bibitem{#1}}    
\nc{\Label}[1]{\label{#1}}    
    
\nc{\Mpc}{Mpc/h}    
\nc{\vev}[1]{\langle #1 \rangle}    
    
\nc{\eea}{\end{eqnarray}} \nc{\ee}{\end{equation}}  
\nc{\eeq}{\end{equation}}  
    
\def\lcdm{$\Lambda$CDM~}

\def\ltsima{$\; \buildrel < \over \sim \;$}    
\def\gtsima{$\; \buildrel > \over \sim \;$}    
\def\simlt{\lower.5ex\hbox{\ltsima}}    
\def\simgt{\lower.5ex\hbox{\gtsima}}    
\nc{\w}{$w_2(\theta)$\ }    
\nc{\ie}{i.e.}     
\nc{\eg}{e.g.}    
\def\q{{\hat n} }

\input epsf

\begin{document}

\title[Cross-correlation between WMAP and the APM survey]{Measurement of the gravitational potential evolution from the cross-correlation between WMAP and the APM Galaxy survey}    
    
\author[Fosalba \& Gazta\~{n}aga]{    
Pablo Fosalba$^{1}$, Enrique Gazta\~{n}aga$^{2,3}$ \\  
$^{1}$Institut d'Astrophysique de Paris, 98bis Bd Arago, 75014 Paris, France \\    
$^{2}$Institut d'Estudis Espacials de Catalunya. IEEC/CSIC, Gran Capit\'{a}n 2-4, 08034 Barcelona, Spain\\    
$^{3}$INAOE, Astrofisica, Tonantzintla, Puebla 7200, Mexico    
}

\maketitle  

\begin{abstract}    
    
Models with late time cosmic acceleration, such as the     
$\Lambda$-dominated CDM model, predict a  freeze out for the  growth    
of linear gravitational potential at moderate redshift $z<1$, what      
can be observed as temperature anisotropies in the CMB: the so called    
integrated Sachs-Wolfe (ISW) effect.     
We present a direct measurement of the ISW     
effect based on the angular cross-correlation function,    
$w_{TG}(\theta)$, of CMB temperature anisotropies     
and dark-matter fluctuations traced by galaxies. We cross-correlate    
the first-year WMAP data in combination with the APM Galaxy    
survey. On the largest scales, $\theta = 4-10$ deg,     
we detect a non-vanishing cross-correlation at 98.8\% significance level,    
with a 1-$\sigma$ error of  $w_{TG} = 0.35 \pm 0.14 \mu$K,     
what favors large values of  $\Omega_\Lambda \simeq 0.8$ for flat FRW models.     
On smaller scales, $\theta < 1$deg,   
the correlations disappear. This is contrary to what would be  
expected from the ISW effect, but the absence of correlations  
may be simply explained if the ISW signal was being cancelled by  
anti-correlations arising from the thermal Sunyaev-Zeldovich (SZ) effect.    
\end{abstract}    
    
    
    
\maketitle    
    
    
\section{Introduction}    
\label{sec:intro}    
    
The recent    
measurements of CMB anisotropies made public by the WMAP team    
are in good agreement with a `concordance'     
cosmology based on the \lcdm model. The    
unprecedented sensitivity, frequency and sky coverage of this new data set    
provides us the opportunity of asking new questions about the    
evolution of the universe.    
In this paper we present the     
cross-correlation of the cosmic microwave background (CMB)    
anisotropies measured by WMAP \cite{B03a}, with     
galaxies in the APM Galaxy Survey \cite{M90}.     
In the observationally favored \lcdm models, a non-vanishing     
CMB-galaxy cross-correlation signal arises from     
the distortion of the pattern of    
primary CMB anisotropies by the large-scale structures as microwave photons    
travel from the last scattering surface to us. On large angular scales    
such distortion is mainly produced by the energy injection photons experience    
as they cross time-evolving dark-matter gravitational potential wells:    
the so-called integrated Sachs-Wolfe effect \cite{SW67}.    
On smaller scales additional secondary anisotropies are produced     
when photons scatter off hot intra-cluster gas, ie     
the Sunyaev-Zeldovich effect \cite{SZ69}.    
    
In this work, we shall use optical galaxies from the APM survey    
as tracers of the large-scale dark-matter distribution of the universe.    
The APM Survey has produced one of the best estimates of the angular galaxy    
2-point correlation function to date.  Its shape on large scales    
led to the discovery of excess large-scale power, and gave early    
indications of the \lcdm model \cite{E90,M90,BE93},     
\cite{G95}.     
Higher-order correlations have also been studied    
in the  APM Galaxy Survey \cite{G94,S95,FG99}. For the    
first time, these measurements were accurate enough and    
extended to sufficiently large scales to    
probe the weakly non-linear regime with a reliable Survey.      
The results are in good agreement with    
gravitational growth for a model with initial Gaussian fluctuations.  They    
also indicate that the APM galaxies are relatively unbiased tracers of the    
mass on large scales \cite{GJ01}.    
Moreover the APM results are in excellent agreement     
with other wide field photometric    
surveys, such as the Sloan Digital Sky Survey (SDSS),     
for both number counts and clustering     
\cite{D01,Sc02,G02a,G02b}.

\section{Data \& simulations}    
\label{sec:data}    
    
The APM Galaxy Survey \cite{M90}    
is based on 185 UK IIIA-J Schmidt photographic plates each corresponding    
to $5.8\times 5.8$ deg$^2$ on the sky limited to $b_J \simeq 20.5$ and    
having a mean depth of $\simeq 400$ Mpc/h for $b <-40$ deg and $\delta<-20$    
deg.  These fields were scanned by the APM machine and carefully matched    
using the $5.8\times 0.8$ deg$^2$ plate overlaps.     
Out of the APM Survey we considered a $17<b_J<20$    
magnitude slice, which includes 1.2 million galaxies at a mean redshift    
$\bar{z} = 0.15$, in an equal-area projection pixel map with a    
resolution of $3.5'$, that covers over $4300$ deg$^2$    
around the SGC.

\begin{figure*}    
{\centering    
{\epsfysize=10cm \epsfbox{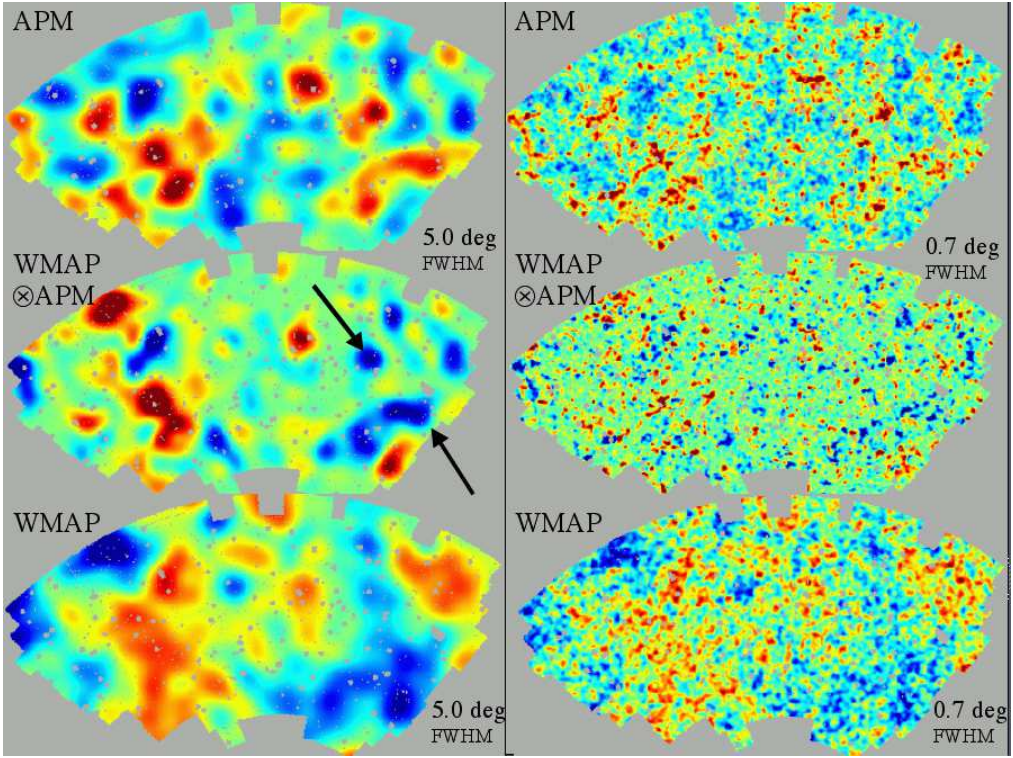}}    
}    
\caption{\label{fig:apmwmap}    
APM galaxy density fluctuation maps (top panels)    
compared to WMAP (V-band) maps (bottom panels) and    
the cross-correlation map (middle). In each case    
left panels show the maps smoothed with a Gaussian beam of FWHM $=5$ deg,    
while right panels have FWHM $=0.7$ deg. We use normalized units (dimensionless)    
with linear color scheme in the range ($-3\sigma,+3\sigma$),     
being $\sigma$ the pixel variance in each map.}    
\end{figure*}

We use the full-sky CMB maps from     
the first-year WMAP data \cite{B03a}.     
In particular, we have chosen the V-band ($\sim 61$ GHz)     
for our analysis since it     
has a lower pixel noise than the highest frequency W-band    
($\sim 94$ GHz), while it has sufficiently high spatial resolution    
($21^{\prime}$) to map    
the typical Abell cluster radius at the mean APM depth.    
We mask out pixels using Kp0 mask, which cuts $21.4 \%$    
of sky pixels \cite{B03b},    
to make sure Galactic emission does not affect our analysis.    
WMAP and APM data are digitized into $7^{\prime}$ pixels using the    
HEALPix tessellation     
\footnote{Some of the results in this paper have been     
derived using HEALPix \cite{GHW99},   
http://www.eso.org/science/healpix }.    
Figs \ref{fig:apmwmap} show these maps    
smoothed using a Gaussian beam of FWHM $= 5$ deg (left) and $0.7$ deg (right    
panels).

\label{sec:null}    
    
To determine the accuracy of our error estimation we run $200$    
WMAP V-band Monte-Carlo realizations. We simulate the    
signal by making random realizations     
of the CMB angular power-spectrum     
as measured by WMAP,   
convolved with its measured symmetrized beam profile,  
to which we add random    
realizations of the white noise estimated for the V-band     
\cite{H03}.     
Sampling variance in the WMAP-APM cross-correlation is thus evaluated    
by computing the correlation between the simulated V-band CMB maps     
(with WMAP Kp0 mask pixels removed) and the APM survey.

    
\section{WMAP-APM Cross-Correlation}    
\label{sec:cross}

\begin{figure}    
\centering{\epsfysize=7.5cm \epsfbox{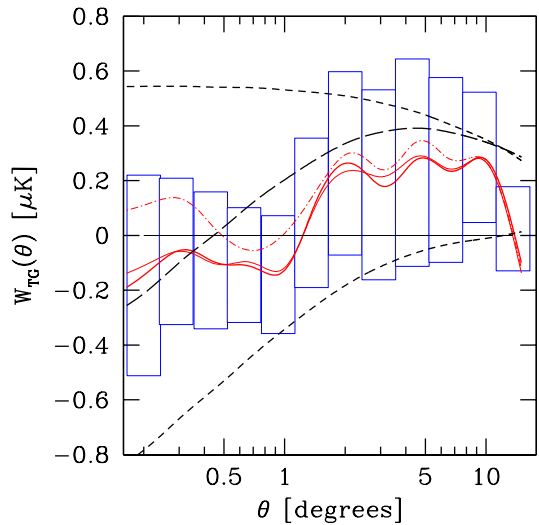}}    
\caption{\label{fig:jacknife}    
Comparing measurement to predictions:    
The two solid lines and the dotted    
line show $w_{TG}$ results for WMAP bands V, W, and the foreground ``cleaned''    
map.  Boxes give the $68\%$ confidence levels.    
Also shown are the theoretical predictions     
for ISW  and SZ (upper and lower short-dashed lines)    
and their sum (long-dashed line) for the best-fit model     
with $\Omega_{\Lambda}\simeq 0.8$.}    
\end{figure}

We define the cross-correlation function as the expectation value of density    
fluctuations $\delta_G= N_G/<N_G>-1$ and temperature anisotropies    
$\Delta_T= T-T_0$ (in $\mu$K)    
at two positions $\q_1$ and $\q_2$ in the sky:    
$w_{TG}(\theta) \equiv  \vev{ \Delta_T({\bf\q_1}) \delta_G({\bf\q_2}) }$,    
where $\theta = |\bf{\q_2}-\bf{\q_1}|$, assuming that    
the distribution is statistically isotropic.    
To estimate $w_{TG}(\theta)$ from the pixel maps we use:    
\be{eqn:ctg}    
w_{TG}(\theta) = {\sum_{i,j} \Delta_T({\bf\q_i})~     
\delta_G({\bf\q_j}) ~w_i~w_j \over{\sum_{i,j} w_i~w_j}},    
\ee    
where the sum extends to all pairs $i,j$ separated by     
$\theta \pm \Delta\theta$.     
The mean temperature fluctuation is subtracted $\vev{\Delta_T}=0$.    
The weights $w_i$ can be used to minimize the variance when the pixel    
noise is not uniform, however this introduces larger cosmic variance.    
Here we follow the WMAP team and use uniform weights (i.e. $w_i=1$).     
We consider angular scales, $\theta < 10$ deg. Cross-correlations are    
expected to be dominated by sampling variance beyond  $\sim    
10$ deg, where the APM angular correlation function vanishes.    
Fig \ref{fig:jacknife} shows the resulting cross-correlation.    
On scales above $\theta > 5$ deg there    
is a significant correlation above the estimated error-bars.

Fig \ref{fig:jacknife} shows the 1-$\sigma$ confidence interval     
for $w_{TG}(\theta)$ obtained using the jack-knife covariance matrix.     
Surveys are first divided into $M=10$ (similar results are found    
for $M=10-20$) separate regions on the sky, each of equal area. The analysis    
is then performed $M$ times, each time removing a different region,     
the so-called jack-knife subsamples, which we label $k=1 \dots M$.    
The estimated statistical covariance for $w_{TG}$ at scales     
$\theta_i$ and $\theta_j$ is then given by:    
\be{covarjack}    
C_{ij}    
= {M-1\over{M}} \sum_{k=1}^{M} \Delta w_{TG}^k(\theta_i) \Delta    
w_{TG}^k(\theta_j)     
\ee    
where $\Delta w_{TG}^k(\theta_i)$ is the difference between the    
$k$-th subsample and the mean value for the $M$ subsamples.    
The case $i=j$ gives the error variance.     
The accuracy of the jack-knife covariance have been tested    
for both WMAP \cite{G03} and the APM and SDSS survey    
\cite{Z02,Sc02,G02a,G02b}.    
 
\begin{figure}    
\centering{    
\vskip -3.5truecm     
\epsfysize=7.5cm \epsfbox{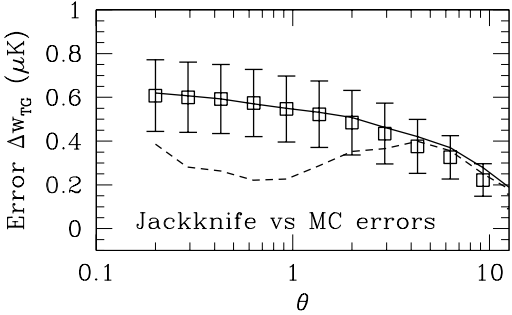}}    
\caption{\label{fig:jackvs}    
Errors in the cross-correlation $w_{TG}(\theta)$    
from the dispersion in 200    
Monte-Carlo simulations (solid line) as compared     
with the mean and dispersion (squares with errobars)    
in the jack-knife error estimation over the same simulations.    
Dashed line correspond to the jack-knife     
error in the real WMAP-APM sample.}    
\end{figure}

We have used the Monte-Carlo (MC) simulations described above
to find that {\it the jack-knife (JK) errors from a single
MC simulation} agree very well (better than  $20\%$ accuracy), with the
MC error from 200 realizations. This is shown in
Fig. \ref{fig:jackvs}. The JK error from a
{\it single realization} (shown as squares) is in excellent agreement with
the MC error (dispersion from 200 realizations, shown as continuous
line). Thus, in our case, using JK errors over a single realization gives
an unbiased estimation of the true error, but the variance (shown
as errorbars in Fig. \ref{fig:jackvs}) in this error estimation
can be as large as $20\%$. 

On the other hand, {\it the JK errors in the real WMAP-APM sample}
are shown as a dashed line in Fig. \ref{fig:jackvs}.
They are comparable to the errors we measure using the MC simulations
on scales $\theta>4$ deg, where the JK errors are potentially subject to
larger biases (as we approach the size of the JK subsample).
On smaller scales, the JK errors from WMAP
are up to a factor 2 smaller than the JK errors
(or sample to sample dispersion) within the MC simulations.
The deviation is significant, given the errorbars, and
it therefore shows that the MC simulations (rather than the JK
error method) fail to reproduce the WMAP-APM data on small scales.
This is not very surprising as the MC simulations do
not include any physical correlation that might be present in WMAP-APM
and assume a CMB power spectrum that is valid for the
whole sky, and not constraint to reproduce the WMAP power over the
APM region. Alternatively, the JK errors provide a model free
estimation that is only subject to moderate ($20\%$) uncertainty,
while MC errors depend crucially on the model assumptions used
to produce simulations.

Despite these differences in the errors on small scales, the overall    
significance for the detection on different    
scales is similar when we use the $\chi^2$ values below or when we ask    
how many of the 200 MC simulations have a signal and JK error    
comparable to the observations.  When a particular MC simulation has    
an accidentally large cross-correlation signal, it also has a large     
noise (JK error) associated.  We find that at 5 degrees only one   
of MC simulations have a signal to noise ratio comparable to   
the observed WMAP-APM correlation on large scales.  
This means that the    
significance of the cross-correlation detection is better than $1\%$.    
We can now estimate confidence regions in $w_{TG}$    
using a $\chi^2$ test with the JK covariance matrix:     
\be{eq:chi}    
\chi^2 = \sum_{i,j=1}^{N} \Delta_i ~ C_{ij}^{-1} ~ \Delta_j,    
\ee    
where $\Delta_i \equiv w_{TG}^E(\theta_i) - w_{TG}^M(\theta_i)$    
is the difference between the "estimation" $E$ and the    
model $M$. The $\chi^2$ test gives a best fit constant    
$w_{TG}= 0.35 \pm 0.13 \mu$K  (the error corresponds to 68\% C.L.  
$\chi^2_{min}=0.19$)    
for the 3 bins in the range $\theta=4-10$ degrees.    
For this  range of scales,  a constant null correlation $w_{TG}= 0$  
gives $\Delta\chi^2=6.1$  which has a probability of $P=1.2\%$ and sets    
the significance of our detection to $2.5 \sigma$.    
We find similar results when we use the MC covariance matrix  
(eg,  $\chi^2_{min}=0.33$ for  
 best fit; $\Delta\chi=5.3$ and $P=2\%$ for the   
significance of the detection) but concentrate on the JK matrix  
 from now on  for the reasons given above.

\subsection{Comparison with Predictions}    
\label{sec:isw}    
    
Galaxy fluctuations in the sky can be modeled as,    
$\delta_G(\q) = \int dz ~\phi_G(z)~ \delta_G(\q,z)$    
where $\phi_G(z)$     
models the survey selection function along the line-of-sight.    
We will assume here that APM    
galaxies are good tracers of the mass on large scales     
(see \S\ref{sec:intro}), so that we can use the linear bias    
relation: $\delta_G \simeq b \delta$, with $b\simeq 1$     
and for the power spectrum: $P_G(k,z) \simeq b(z)^2 P(k,z)$.     
In the linear regime we further have: $P(k,z)= D^2(z)~P(k)$.    
We can then define a galaxy window function $W_G(z) \simeq b(z)~D(z)~\phi_G(z)$    
accounting for bias, linear growth, and the galaxy selection function.    
For the APM selection we use the function in \cite{BE93}     
with a mean redshift $\bar{z} \simeq 0.15$.    
Thus the galaxy 2-point angular correlation    
is \cite{BE93},\cite{GB98}    
$w_{GG}(\theta)  =<\delta_G \delta_G> = \int dk~k~P(k)~g_G(k\theta)$,    
where the kernel $g_G(k\,\theta)$ is a line-of-sight integral,    
$g_G(k\theta) = {1/{2\pi}}~\int dz~W_G^2(z)~j_0(k\theta\,r)$    
where $j_0$ is the zero-th order Bessel function,    
and $r(z)$ denotes comoving distance.    
We use \cite{B84} for the linear power spectrum, with shape parameter     
$\Gamma = \Omega_m h$,       
and $h=0.7$. We take $\sigma_8=1$    
and $b=1$ which give a resonable match to the measured variance in    
the APM  \cite{G95}.

The temperature of CMB photons is gravitationally     
blueshifted as they travel through the time-evolving   
dark-matter gravitational    
potential wells along the line-of-sight, from the last scattering    
surface $z_s = 1089$    
to us today, $z=0$ \cite{SW67}.    
At a given sky position ${\bf\q}$:    
$\Delta T^{ISW}({\bf\q}) = -2 \int ~dz~\dot{\Phi}({\bf\q},z)$.    
For a flat universe $\nabla^2\Phi = -4\pi G a^2 \rho_m \delta$     
\cite{P80}, which in Fourier space reads,    
$\Phi(k,z) = -3/2 \Omega_m ~(H_0/k)^2~\delta(k,z)/a$. Thus:    
\bea{eq:wTG_ISW}    
w_{TG}^{ISW}(\theta) &=& <\Delta_T^{ISW}\delta_G> = \int {dk\over k}~P(k)~g(k\theta) \\    
g(k\theta) &=& {1\over{2\pi}}~\int dz~W_{ISW}(z)~W_G(z)~j_0(k\theta\,r) \nonumber    
\eea    
where the ISW window function is given by     
$W_{ISW} = -3~\Omega_m~({H_0\over c})^2~\dot{F}(z)$, with    
$c/H_0 \simeq 3000 h$ Mpc$^{-1}$ the Hubble radius today.    
$\dot{F} = d(D/a)/dr = (H/c) D (f-1)$ with $f \simeq \Omega_m^{6/11}(z)$    
quantifies the time evolution of the gravitational potential.    
Note that $\dot{F}$ decreases as a function of increasing redshift    
(as $\Omega_m(z)  \rightarrow 1$).     
It turns out that for flat universes,    
$\Omega_m+\Omega_\Lambda=1$,      
$W_{ISW}$ has a maximum at $\Omega_\Lambda \simeq 0.6$    
and tends to zero both for $\Omega_\Lambda \rightarrow 1$     
(since the prefactor $\Omega_m  \rightarrow 0$)    
and also for  $\Omega_\Lambda \rightarrow 0$     
(because  $\dot{F} \rightarrow 0$).    
The \lcdm prediction with $\Omega_\Lambda \simeq 0.8$ and $\Gamma = 0.14$    
is shown as a short-dashed line in Fig \ref{fig:jacknife}.

\label{sec:sz}

For the thermal Sunyaev-Zeldovich (SZ) effect,     
we just assume that the gas pressure $\delta_{gas}$ fluctuations    
are traced by the APM galaxy     
fluctuations $\delta_{gas} \simeq b_{gas}~\delta_G$     
with an amplitude $b_{gas} \simeq 2$, representative    
of galaxy clusters. Note that analytical results based on halo models     
and hydrodynamic simulations show that this ``gas bias''    
factor is scale and redshift dependent \cite{RT02}. However, for low-$z$    
sources and {\it linear} scales one can safely take $b_{gas} = 2-4$.  
Note that the cross-correlation function is dominated  
(its overall shape and amplitude) by large-scale modes that  
are well described by linear theory, but a more precise calculation  
requieres non-linear corrections.  
Thus a rough conservative estimate is given by \cite{R00}:    
$w_{TG}^{SZ}(\theta) = -b_{gas}~\overline{{\Delta T}} w_{GG}(\theta)$,    
where $\overline{{\Delta T}} \simeq 6.6 \mu$K is the estimated mean     
SZ fluctuation from APM clusters, what corresponds to a Compton    
parameter $y \simeq 2\times 10^{-7}$ \cite{R00}.     
The SZ prediction along with the total correlation,    
$w_{TG}=w_{TG}^{ISW}+w_{TG}^{SZ}$, are given by the lower short-dashed and    
long-dashed lines in Fig \ref{fig:jacknife}. Deriving more accurate parameter 
contraints from the SZ effect requires including non-linear effects in the 
power spectrum, what is beyond the scope of this paper.

\section{Discussion}    
\label{sec:discuss}    
    
The main result of this paper is a measurement of a positive    
cross-correlation $w_{TG} = 0.35 \pm 0.13 \mu$K  (1-$\sigma$ error)     
between WMAP CMB temperature anisotropies and the    
Galaxy density fluctuations in the largest scales of the APM galaxy    
survey, $\theta \simeq 4-10$ deg. The significance of this    
detection is at the $98.8\%$ confidence level (2.5 $\sigma$).    
Large-scale modes from the primary     
SW temperature anisotropies     
introduce large sampling variance and  makes     
measurements of the ISW contribution intrinsically noisy.     
The measured cross-correlation on $\theta > 4$ deg scales     
is in good agreement with ISW effect predictions    
from standard $\Lambda$CDM models. Using the theoretical modeling    
in \S III.A we find a 2-$\sigma$ interval of $\Omega_\Lambda=0.53-0.86$,    
with a best fit value of $\Omega_\Lambda \simeq 0.8$.

If the detected cross-correlation is only due to the ISW    
effect  \cite{CT96},    
one would expect a stronger ISW-induced correlation on smaller scales     
(see Fig \ref{fig:jacknife}). Instead,    
on scales $0.2 < \theta < 1$ deg, the mean cross-correlation becomes    
negative, $w_{TG} \approx -0.06\pm 0.16 \mu$K.     
This sudden drop can be understood as    
thermal SZ contribution from hot gas in galaxy    
clusters \cite{R00}.     
The SZ effect contributes to a level    
$w_{TG}^{SZ}= w_{TG}-w_{TG}^{ISW} \approx -0.41\pm 0.13 \pm 0.16\mu$K, where    
 errors reflect the  uncertainties at large and small scales.    
This result can be used to set bounds on the mean Compton scattering     
of CMB photons crossing clusters  (see also \cite{D03}).    
Despite the high galactic latitude  ($b <-40$ deg),     
our results can potentially be contaminated by     
Galactic dust \cite{N03}.     
However, as illustrated in Fig 2, using the Kp0 masked W-band, V-band or    
a foreground ``cleaned''map \cite{T03} all give similar    
results within the errors.

Fig \ref{fig:apmwmap} shows    
the ISW and SZ contributions at the map level.    
On larger scales the APM-WMAP product map shows a clear correlation with the    
APM structures, while on smaller scales (right panels) this correlation    
fades away or turns into anti-correlation at the core of    
the largest structures in the APM map. These APM structures    
correspond to the very large scale potentials hosting    
superclusters or a few large clusters in projection.     
Some of them appear to be anti-correlated with WMAP in the    
product map, but with very similar shapes    
(regions pointed by an arrow in Fig \ref{fig:apmwmap}).    
This is not surprising since the measurement of the ISW effect     
is intrinsically affected by sampling variance from the larger    
amplitude modes due to primordial SW fluctuations.    
In particular, if a real (positive amplitude)    
ISW signal is ``mounted'' over a larger scale SW mode     
(of negative amplitude) it can produce a net negative contribution to $w_{TG}$.    
    
Our findings are in agreement with    
recent work on the cross-correlation measure of WMAP    
with NRAO VLA Sky Survey radio source catalogue (NVSS)    
\cite{BC04}, \cite{N03}.    
They detect a signal of $w_{TG} \simeq 0.16 \mu$K with $1.8$ deg pixels    
(148 counts/pixel), what is consistent with     
our measurements once the different selection function is taken into account.    
Our cross-correlation analysis on large-scales    
reveals that the evolution of the gravitational potential has     
been strongly suppressed at low-$z$. The drop of this positive     
correlation on small-scales suggests that we might be measuring    
the SZ-induced distortion of CMB photons by nearby clusters.     
Deep large area galaxy surveys, such as the SDSS, should be able    
to confirm these results, provide tighter constraints on     
cosmological parameters and improve our knowledge of cluster physics    
\cite{PS00}. Such analysis, together with a detailed     
treatment of the SZ and lensing effects    
will be presented elsewhere \cite{F03}.    
    
\section*{Acknowledgments}  
    
We thank F.Castander for useful discussions.    
We acknowledge support from the Barcelona-Paris bilateral project    
(Picasso Programme).    
PF acknowledges a post-doctoral CMBNet fellowship from the European Commission.    
EG acknowledged support from INAOE, the Spanish Ministerio de Ciencia y    
Tecnologia, project AYA2002-00850, EC-FEDER funding.

\end{document}